\documentclass[aps,twocolumn,superscriptaddress,floatfix,longbibliography]{revtex4-1}
\usepackage{dcolumn}
\usepackage{amsmath}
\usepackage{amssymb}
\usepackage{amsthm}
\usepackage{graphicx}
\usepackage{bm}
\usepackage[T1]{fontenc}
\usepackage{color}
\usepackage{url}
\usepackage[bf]{subfigure}
\usepackage{rotating}
\usepackage{scalefnt}
\usepackage{multirow}

\usepackage{scalefnt}
\usepackage{times}
\usepackage{mathtools}
\usepackage{bbm}

\usepackage{float}
\usepackage{enumerate}

\usepackage{nicefrac}

\makeindex

\begin{document}

\title{On the temporal resolution limits of numerical simulations in complex systems}

\author{Guilherme Ferraz de Arruda}
\affiliation{CENTAI Institute, Turin, Italy}

\author{Yamir Moreno}
\affiliation{Institute for Biocomputation and Physics of Complex Systems (BIFI), University of Zaragoza, Zaragoza 50009, Spain}
\affiliation{Department of Theoretical Physics, University of Zaragoza, Zaragoza 50009, Spain}
\affiliation{CENTAI Institute, Turin, Italy}

\begin{abstract}
In this paper we formalize, using the Nyquist-Shannon theorem, a fundamental temporal resolution limit for numerical experiments in complex systems. A consequence of this limit is aliasing, the introduction of spurious frequencies due to sampling. By imposing these limits on the uncertainty principle in harmonic analysis, we show that by increasing the sampling interval $\Delta t$, we can also artificially stretch the temporal behavior of our numerical experiment. Importantly, in limiting cases, we could even observe a new artificially created absorbing state. Our findings are validated in deterministic and stochastic simulations. In deterministic systems, we analyzed the Kuramoto model in which aliasing could be observed. In stochastic simulations, we formalized and compared different simulation approaches and showed their temporal limits. Gillespie-like simulations fully capture the continuous-time Markov chain processes, being lossless. Asynchronous cellular automata methods capture the same transitions as the continuous-time process but lose the temporal information about the process. Finally, synchronous cellular automata simulations solve a sampled chain. By comparing these methods, we show that if $\Delta t$ is not small enough, the cellular automata approach fails to capture the original continuous-time Markov chain since the sampling is already built into the simulation method. Our results point to a fundamental limitation that cannot be overcome by traditional methods of numerical simulations.
\end{abstract}

\maketitle

Sampling is often an unavoidable consequence of most computational methods. This is the case for many systems, ranging from digital signal processing, where this issue is well studied, to numerical experiments in general, where it remains underexplored. In particular, we may have higher frequencies in a networked system due to the interaction of two different signals. For example, consider a process in which a node outputs a nonlinear function of several input signals arriving via its incoming edges. This output can result from a convolution of the signals in the frequency domain, allowing the output frequencies to be higher than the input frequencies. Remarkably, this effect can be amplified for a signal traveling in a network where the output of one node is the input of another. This setup can even create positive or negative feedback due to possible loops in the structure. Taking this factor into account is crucial if one is interested in understanding or predicting temporal quantities in real systems. This work shows that neglecting this issue can lead to inaccurate or incorrect conclusions and provides insights on how to adequately account for this effect in the analysis of complex systems.

A well-known limit is formalized by the Nyquist-Shannon theorem, which states that \emph{``if a function $x(t)$ contains no frequencies higher than $f_{\max}$ Hertz, it is completely determined by giving its ordinates at a series of points $\nicefrac{1}{(2 f_{\max})}$ seconds apart''}~\cite{Shannon1949}. In this case, the so-called Nyquist frequency is $f_N = 2 f_{\max}$. Consequently, when sampling below the Nyquist-Shannon frequency, the perceived frequency depends on the sampling frequency $f_s$~\cite{oppenheim1997signals}.
So, the perceived frequency is
\begin{equation}
 f_{perceived} = \big | f_{\max} - f_s \times NINT \left( \nicefrac{f}{f_s} \right) \big |,
\end{equation}
where $NINT \left( \cdot \right)$ is the ``nearest integer'' function.
Although this is a well-known result in communications, it is often neglected in numerical experiments.

Here we explore two non-extensive categories of numerical experiments: (1) deterministic and (2) stochastic (Monte Carlo) simulations. The second is further divided according to temporal approaches: (i) continuous-time and (ii) discrete-time cellular automata: (ii.a) asynchronous and (ii.b) synchronous. In deterministic cases, the initial conditions or parameters may be random, but the temporal trajectories are defined once these choices are made. In the stochastic scenario, a Monte Carlo simulation represents a trajectory (sequence of microstates) in the space of possible trajectories. In this case, we are often interested in the mean and variance of these trajectories given the same initial condition.

In practice, a deterministic process on top of a network can be described by a function of the form $\dfrac{d x_i}{dt} = f(\bm x, \bm \lambda, t)$, where $\bm x = \left[x_1, x_2, \cdots, x_N \right]$ is the state vector, $\bm \lambda$ are the parameters of the model, and $t$ encodes the temporal dependency. For simplicity, we focus on processes in which a single state is accounted for. However, our argument can be easily extended to more dimensions.
Often we do not have an analytical solution for the model. In this case, we rely on the numerical solution of the above-defined ODE.
Despite the numerical errors the numerical methods make, we implicitly sample our signal every $\Delta t$.
Thus, the maximum observed frequency will be $f_N = \nicefrac{1}{(2 \Delta t)}$.
We highlight that even mean-field approximations might be affected in this case.

\begin{figure*}[t]
   \includegraphics[width=\textwidth]{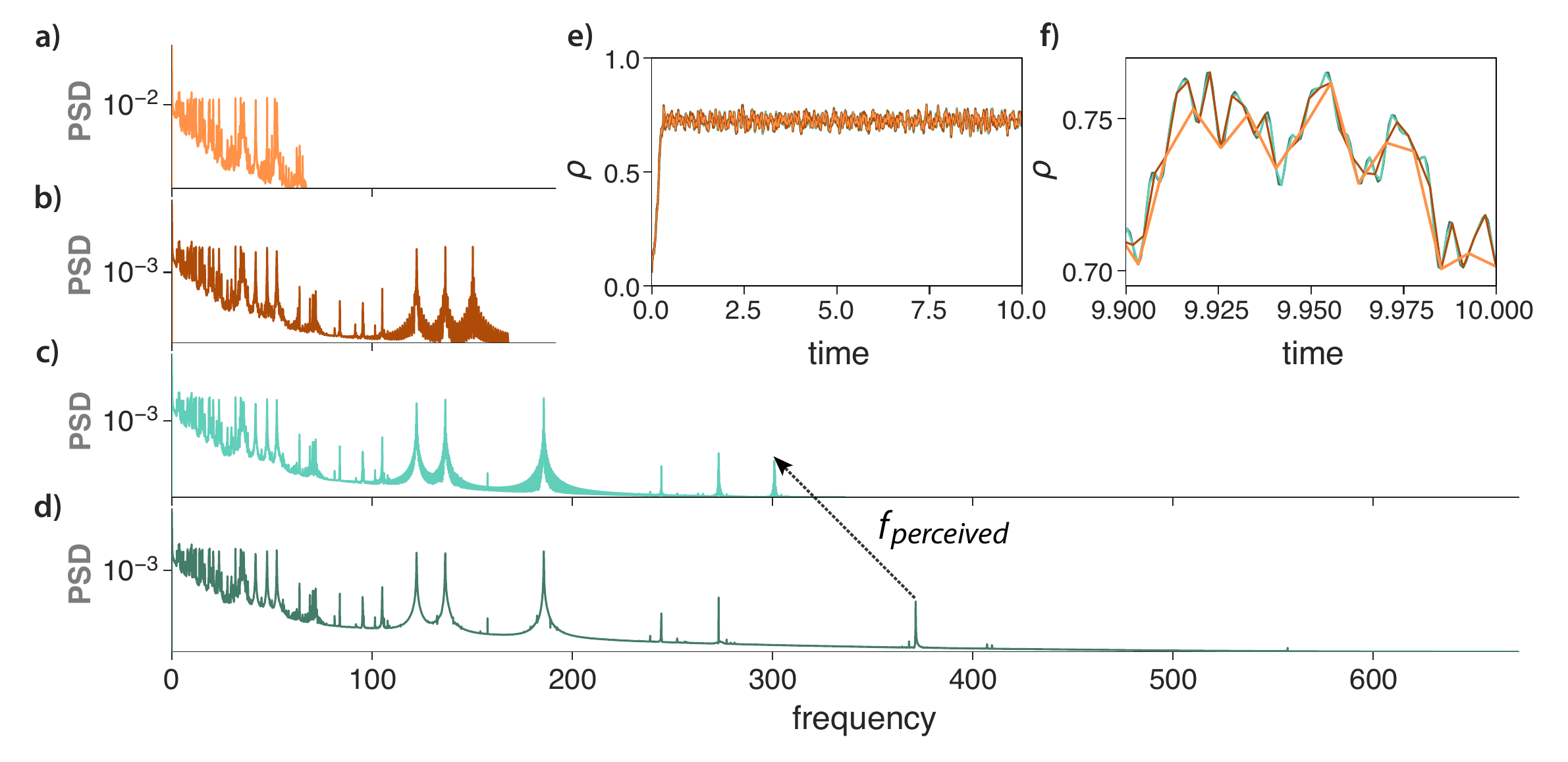}
   \caption{In (a) through (d), we have the power spectrum density (PSD) for $\Delta t = \nicefrac{1}{134.54}  \approx \nicefrac{5}{2\max_i \omega_i}$, $\Delta t = \nicefrac{1}{336.35} \approx \nicefrac{1}{\max_i \omega_i}$, $\Delta t = \nicefrac{1}{672.7} \approx \nicefrac{1}{2\max_i \omega_i}$, $\Delta t = \nicefrac{1}{1345.4} \approx \nicefrac{1}{4\max_i \omega_i}$, respectively. We have sampling frequencies below and above the estimated Nyquist frequency ($f_N^{\text{est}} = 2 \max_i \omega_i \approx 672.7$), where the estimation is based on the natural frequencies. In (e) and (f), we show the temporal behavior of the order parameter for all $\Delta t$ values. In (f), we focus on the interval $[9.9, 10]$ and show the effect of each sampling. Here, we consider an Erd\"os -- Renyi network with $N = 250$ nodes, $p = 0.05$, coupling strength $\lambda = 4.5$, and we sampled the natural frequencies from a Lorentz distribution with a scale parameter $\gamma = 10$ for visualization purposes.}
   \label{fig:Kuramoto}
\end{figure*}

As an example, we analyze the Kuramoto model~\cite{Rodrigues2016} (see SM). To understand the effects of sampling, we solve it numerically using the LSODA algorithm~\cite{hindmarsh1982, Petzold1983, 2020SciPy-NMeth} with a small $\Delta t$ ($\Delta t = \nicefrac{1}{1345.4} \approx \nicefrac{1}{4\max_i \omega_i}$, where $\omega_i$'s are the natural frequencies) and analyze the power spectrum density for different samples from that solution. Solving the system with different $\Delta t$'s would yield similar results. In our example, we extracted the natural frequencies from a Lorentz distribution with a scale parameter $\gamma = 10$ and a coupling $\lambda = 4.5$. Both parameters were chosen for visualization purposes. Fig.~\ref{fig:Kuramoto}(a) to (d) show the average power spectrum density (PSD) over all $\sin \left( \theta_i \right)$ signals. The PSD is estimated using the periodogram~\cite{Welch1967, 2020SciPy-NMeth} with the sampling frequency $f_s = \nicefrac{1}{\Delta t}$. In (e) and (f), we show the time behavior of the order parameter, $\rho = \langle e^{\mathbf{i} \theta_i} \rangle$, which measures how synchronous the oscillators are. We notice that the power spectrum (see Fig.~\ref{fig:Kuramoto} (a) to (d)) does not capture the full dynamics when the sampling frequency is small ($f_s \leq f_N$). Most importantly, the maximum perceived frequency can also be misleading if it is below the Nyquist frequency, $f_N = 2f_{\max}$, (see Fig.~\ref{fig:Kuramoto} (c) and (d)).

On the other hand, the consequence of undersampling can be observed in the temporal behavior (see Fig.~\ref{fig:Kuramoto} (f)). Note that the blue curve has the highest sampling frequency, probably above $f_N$. So, we expect the whole signal to be reproduced. Conversely, the sampling frequency in the orange curve is below $f_N$, so it roughly represents the richness observed in the blue curve. In other words, the sampling frequency in the orange curve neglects higher frequency dynamics. In practice, lower sampling frequencies are sufficient if only the mean order parameter is of interest since the mean has zero frequency. This is probably the case in many applications in physics. However, when one is interested in temporal behavior, the sampling frequency plays a significant role in the analysis. More importantly, we point out a very general problem that can be present in any deterministic model, including, for example, neural dynamics, some opinion models, and even population models.

Next, consider stochastic dynamics. Assuming we have $N$ agents, each agent can be in one of $K$ mutually exclusive states. Thus, this process will have $M = K^N$ possible states. Here, we use the letters $i$ and $j$ to index all possible microstates. We also define the microstate $i$ by the vector $\bm x_i$, whose components $[\bm x_i]_k \in \{ 0, 1,\cdots, K-1 \}$ denote the state of node $k$. Furthermore, $\bm s(t) = \left[ s_0(t), ... s_{M-1}(t) \right]$ accounts for the probability of having each microstate at time $t$. We also restrict ourselves to continuous-time Markov chain (CTMC) processes, which are memoryless and can be formalized by the infinitesimal generator $Q$. The element $Q_{ij}$ is the rate at which the process changes from microstate $i$ to microstate $j$. This stochastic process can be described by a set of $K^M$ linear ordinary differential equations~\cite{Mieghem2009, VanMieghem2014, Mieghem2016, Arruda2018} (see also SM for a full discussion). Thus, its solution will be a summation of exponential functions $e^{\mu_\ell t}$, where $\mu_\ell$ are the eigenvalues of $Q$. It is also instructive to note that the magnitude of the Fourier transform of an exponential $e^{-at} u(t)$ is of the form $\frac{1}{\sqrt{a^2 + \omega^2}}$, where $u(t)$ is the Heaviside function. Due to linearity, the transform for the entire process will be a sum of the individual transforms.

The Fourier transform suggests that to capture all frequencies on the dynamics, we should consider all eigenvalues of $Q$. The multiplicity of $\mu_\ell = 0$ is the number of absorbing states in the process, and using the Gershgorin circle theorem, we can bound the maximum eigenvalue of $Q$ as $\lambda_{max} (Q) \leq \Delta (Q) = 2 \times \max_i \left( \sum_{j \neq i} Q_{ij} \right)$, which is twice the maximum transition rate in the infinitesimal generator. Thus, to be consistent with the Nyquist-Shannon theorem, the sampling rate should be greater than $\Delta (Q)$ (see also SM). Interestingly, we can reach the same conclusion by considering that the frequency of this signal is given by the frequency of the events, i.e., the rates of the Poisson processes. In this case, due to the superposition theorem, the maximum possible frequency for the next event is given by the sum of all frequencies, i.e., $f_{\max} = \max_i \left( \sum_{j \neq i} Q_{ij} \right)$ and $f_N = 2 f_{\max}$.

\begin{figure*}[t]
   \includegraphics[width=\textwidth]{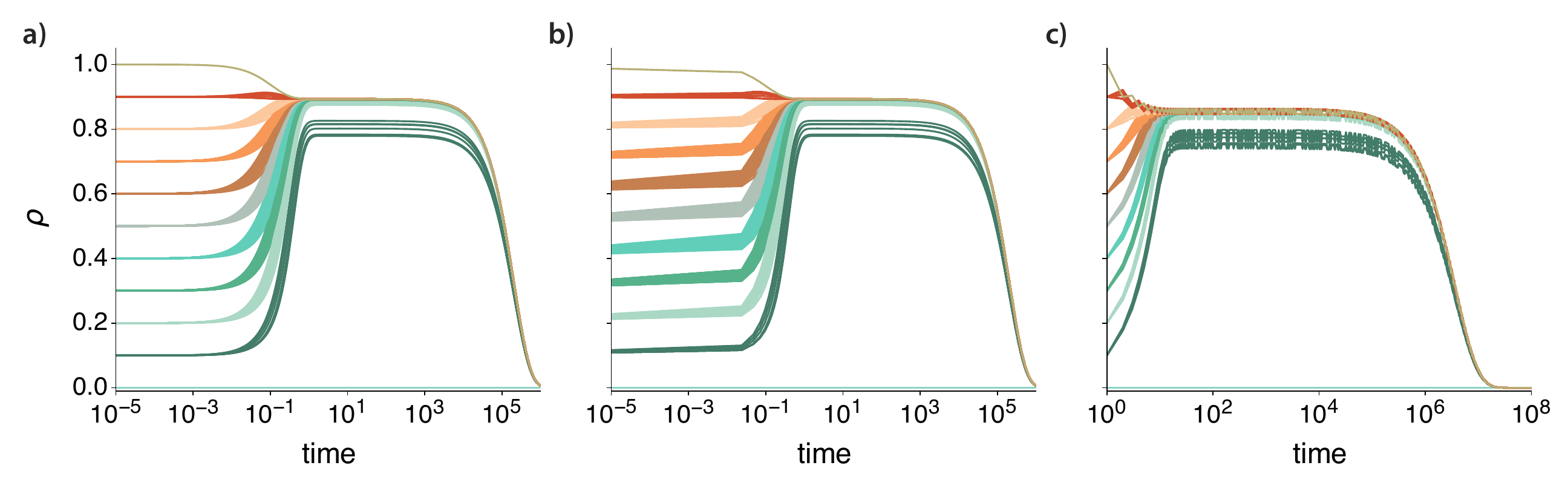}
   \caption{In (a) the temporal evolution of all possible initial conditions (micro-configurations) of the exact solution of the SIS model, in (b) its sampled version, considering only the first-order terms given by $P^S$, with $\Delta t = \nicefrac{1}{\max_i Q_{ii}}$ (limiting case), and the embedded case in (c). Here, we considered $\beta = 2$, $\delta = 1$ on a random network with $N = 10$ nodes and average degree $\langle k \rangle = 5$.}
   \label{fig:Temporal_SIS}
\end{figure*}

From the infinitesimal generator, we can also obtain the probability transition matrix (see also Eq. 10.13 in~\cite{VanMieghem2014}),
\begin{equation}
 P(t) = e^{Qt} = I +\sum_{n=1}^\infty \frac{(Q t)^n}{n!},
\end{equation}
which is continuous for all $t > 0$ (see Ref.~\cite{VanMieghem2014} Lemma 10.2.1).
One can sample the CTMC by obtaining $P^S = P(\Delta t)$ and truncating the sum at $n=1$, thus neglecting terms of $O \left( (\Delta t)^2 \right)$. In this case, if $\Delta t \rightarrow 0$, we recover the CTMC. Note that $\dfrac{d e^{Qt}}{dt} = \lim_{h \rightarrow 0} \frac{P(t+h) - P(t)}{h} = QP(t)$ (see Ref.~\cite{VanMieghem2014}, Lemma 10.2.2). The dynamics described by $s^T(t+\Delta t) = s^T(t) P^S$ is a discrete-time Markov chain (DTMC). Here, time moves in $\Delta t$ steps, and we implicitly assume that multiple events can occur simultaneously. This sampling method is restricted to $\Delta t < \nicefrac{1}{\max_i |Q_{ii}|}$, since larger values of $\Delta t$ imply negative elements in $P^S$. Note that $\nicefrac{1}{\max_i |Q_{ii}|}$ is the minimum expected time for the next event in our dynamics. This guarantees that, on average, each step represents a transition. We note that the smaller $\Delta t$, the more concentrated the diagonal $P^S(\Delta t)$ will be, which implies staying in the same microstate. To our knowledge, this DTMC has no direct Monte Carlo implementation.

Another sampling approach would be to compute the probability of each event (Poisson process) occurring after an arbitrary $\Delta t$. In this case, the transition between the two microstates is the product of the probabilities of change in each node state. Formally, the transition probability matrix is expressed as
\begin{equation}
   \label{eq:P_CA}
   P^{CA}_{ij} = \prod_{k=1}^{N} f_k(\bm x_i, \bm x_j),
\end{equation}
where $f_k(\bm x_i(t), \bm x_j(t))$ is a function that determines the probability that node $k$, in state $[\bm x_i]_k$, transitions to state $[\bm x_j]_k$ after a $\Delta t$ step. The product takes into account the probabilities of all events occurring within the same time window. If $\Delta t$ is small enough, this approach tends to the matrix $P^S$ (see SM for a numerical example). This is because two Poisson processes cannot occur simultaneously (zero probability)~\cite{VanMieghem2014}. 
The dynamics described by Eq.~\eqref{eq:P_CA} represent the synchronous cellular automata (CA) approach to simulations. In this case, we often have local rules that determine the behavior of the agents. The simulation is a process where these rules are sequentially iterated with a fixed time increment. The synchronous cellular automata solve the sampled chain, DTMC, given by the matrix $P^{CA}(\Delta t)$. If $\Delta t$ is small enough ($\Delta t \rightarrow 0$), it should capture the CTMC. However, if $\Delta t$ is larger, the cellular automata approach is not expected to capture the original CTMC accurately. Note that the simulation is limited because it is already sampling the process.

Another discretization approach is the embedded Markov chain (EMC), where we preserve the transitions while losing the temporal information. The probability transition matrix describing this process is
\begin{equation}
 P^{EMC} = I - \left( \text{diag} (Q) \right)^{-1} Q.
\end{equation}
The time spent in each microstate is fixed and equal to one unit of time. Note that a single component on the microstate vector changes at each time step.
The EMC represents asynchronous cellular automata simulations. In this method, a node is chosen at random, and the local rules are applied only to that agent. Then, time is incremented, and the process is repeated.

\begin{figure*}[t]
   \includegraphics[width=\textwidth]{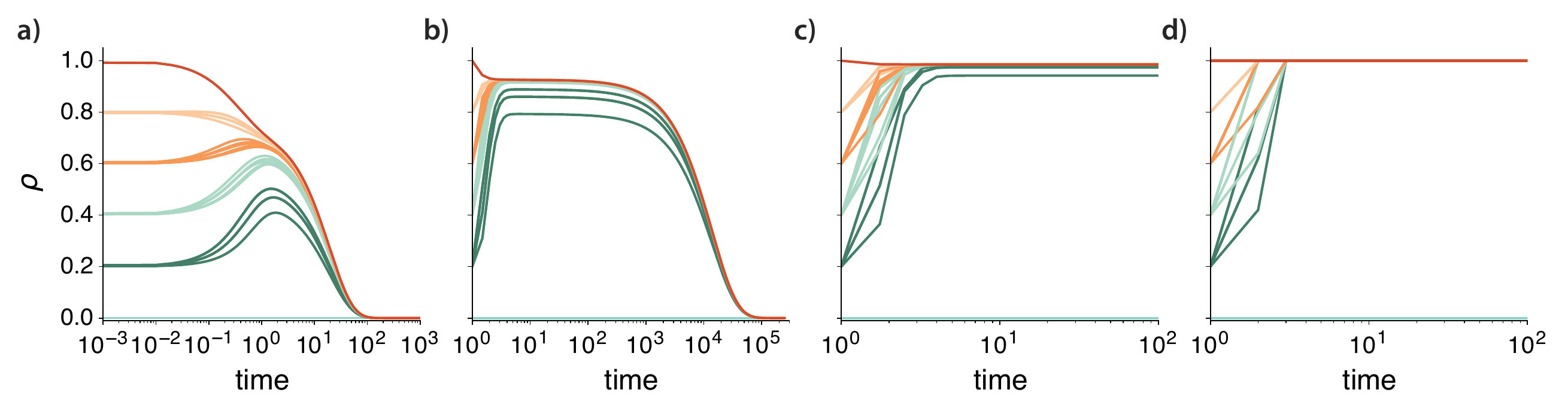}
   \caption{The effect of the sampling frequency on the temporal behavior of the SIS model for several initial conditions. Here, we considered $\beta = 1$, $\delta = 0.9$ on a random network with $N = 5$ nodes and average degree $\langle k \rangle = 3$, and from (a) to (d), we considered $\Delta t = 0.01, 0.5, 0.75, 1$.}
   \label{fig:Sampling}
\end{figure*}

In the following, we provide an example to illustrate the stochastic simulation cases. We consider the susceptible-infected-susceptible (SIS) epidemic spreading model because it is simple enough to be solved exactly for small systems. The contacts are encoded in a graph whose adjacency matrix is denoted $A$. The model is defined as a set of independent Poisson processes. To model the spreading, we associate a Poisson process with rate $\beta$ to each edge from an infected to a susceptible agent. Also, to model recovery, Poisson processes with rate $\delta$ are associated with infected nodes. Here, we follow the exact formulation on networks using the infinitesimal generator in~\cite{Mieghem2009, VanMieghem2014, Arruda2018}.
The infinitesimal generator and the CA transition probabilities are derived and shown in the SM.

Fig.~\ref{fig:Temporal_SIS} shows an example of the CTMC, the sampled case considering $P^S(\Delta t)$ with $\Delta t = \nicefrac{1}{\max_i |Q_{ii}|}$ and the EMC in (a) to (c) respectively (see SM for the numerical methods). This figure emphasizes the temporal limits of $P^S$, where we lose resolution for small $t$ (compare Fig.~\ref{fig:Temporal_SIS} (a) and (b)).
On the other hand, it shows that the EMC not only loses temporal information (note the time each approach takes to reach the absorbing state, i.e., $\rho = 0$) but also creates small oscillations that are not present in the original model.
See also SM for a graphical example of the transition matrices. In addition, in Fig.~\ref{fig:Sampling}, we evaluate the cellular automata approach, $P^{CA}(\Delta t)$, for different values of $\Delta t$.
First, Fig.~\ref{fig:Sampling} (a) is very similar to the CTMC (figure not shown).
As we increase $\Delta t$, we see that the model overestimates the active states. This is evidenced by the time it takes the system to reach the absorbing state.
Although not shown in Fig.~\ref{fig:Sampling}(c), the active states will reach the absorbing state ($\rho = 0$), but it can take a very long time. Importantly, $\rho = 1$ is an absorbing state of the DTMC described by $P^{CA}(\Delta t = 1)$, but not of the CTMC (see Fig.~\ref{eq:P_CA} (d)). We can guarantee that it is an absorbing state only for $\Delta t = 1$ because it is necessary to create a sink on the graph of the transitions (see SM for an example).

The uncertainty principle in harmonic analysis helps us understand the overestimation of the active state. This principle can be described as \emph{``the shorter-lived a function, the wider the band of frequencies given by its Fourier transform; the narrower the band of frequencies of its Fourier transform, the more the function is spread out in time''}~\cite{Hubbard1996}. Formally, this principle is expressed by the inequality~\cite{Pinsky2008}
\begin{equation}
 \left( \frac{\int_{-\infty}^{\infty} t^2 |f(t)|^2 dt}{\int_{-\infty}^{\infty} |f(f)|^2 dt} \right)
 \left( \frac{\int_{-\infty}^{\infty} \xi^2 |\hat{f}(\omega)|^2 d\omega}{\int_{-\infty}^{\infty} |\hat{f}(\omega)|^2 d\omega} \right)
 \geq \frac{1}{4},
\end{equation}
where $f(t)$ is a function over time and $\hat{f}(\omega)$ is its Fourier transform. Given the limits imposed by the Nyquit-Shannnon theorem, we have a limited frequency bandwidth. Thus, the integral in the frequency domain, the second term in the l.h.s., should be in the range $[-f_{\max}, +f_{\max}]$ instead of $(-\infty, +\infty)$, i.e., it will be less than or equal to the original. Consequently, the time signal may have a larger variance over time to account for this artificial frequency constraint. In practice, this can lead to an increase in the time variance, as in our example. Coincidentally, this is the same inequality as in the Heisenberg Uncertainty Principle, but with a slightly different interpretation.

In summary, we have explored the temporal limits of numerical experiments in complex systems.
We used the Kuramoto model to illustrate this problem in deterministic processes. We showed that undersampling can introduce spurious frequencies not present in the original signal. Fortunately, this is not a critical issue if one is only interested in the mean. However, if one is interested in the frequency composition of the signal, the sampling frequency must respect the Nyquist frequency.

Considering stochastic Markovian dynamics, asynchronous cellular automata simulations have a strong and clear limitation regarding their temporal precision since they do not aim to characterize this behavior.
Thus, the relevant comparison is between Gillespie-like simulations and synchronous cellular automata.
From the perspective of the Shannon-Nyquist theorem, we find a significant limitation. Our process is solved as a composition of exponential functions. However, the Fourier transform of an exponential function spans the entire frequency spectrum. Thus, we will have some limitations in reconstructing the signal for any finite sampling frequency.
The highest frequency event in the process has $f_{\max} = \max_i |Q_{ii}|$. So a reasonable estimate would be $\Delta t < \nicefrac{1}{(2\max_i |Q_{ii}|)}$.
In this case, the probability of two events occurring in the same time window is small.
There may also be a practical problem with the increased computational cost. However, this issue is outside the scope of our analysis since cellular automata can be easily parallelized, while for Gillespie methods, this is more challenging.

Although we have focused on a few examples, our results apply to other processes and could even be applied to data-driven approaches where data may be the limiting factor. For example, we could use simulations to estimate missed frequencies due to their sampling frequency. To do this, however, we must have a $\Delta t$ in our simulations that at least respects the Nyquist criteria imposed by the data collection process. From a practitioner's point of view, a significant limitation imposed by the Shannon-Nyquist theorem is that if a simulation is run with a given $\Delta t$, it will only capture processes with a period of $2 \Delta t$. For example, if we set $\Delta t = 1$ day, we will only capture events with a period of two or more days. We hope the ideas presented here provide new perspectives on formalizing numerical experiments in complex systems.

\begin{acknowledgments}
The authors thank Ginestra Bianconi, Henrique Ferraz de Arruda, and Pietro Traversa for fruitful discussions and feedback.
Y.M was partially supported by the Government of Arag\'on, Spain, and ``ERDF A way of making Europe'' through grant E36‐23R (FENOL), and by Ministerio de Ciencia e Innovaci\'on, Agencia Espa\~nola de Investigaci\'on (MCIN/AEI/10.13039/501100011033) Grant No. PID2020‐115800GB‐I00. We acknowledge the use of the computational resources of COSNET Lab at Institute BIFI, funded by Banco Santander (grant Santander‐UZ 2020/0274) and by the Government of Arag\'on (grant UZ-164255). The funders had no role in study design, data collection and analysis, the decision to publish, or the preparation of the manuscript.
\end{acknowledgments}

\appendix

\section{Kuramoto model}
\label{sec:kuramoto}

The Kuramoto model~\cite{Rodrigues2016}, is defined as
\begin{equation}
   \label{eq:Kuramoto}
   \dfrac{d \theta_i}{dt} = \omega_i + \lambda \sum_{j=1}^N A_{ij} \sin(\theta_j - \theta_i),
\end{equation}
where $\theta_i$ and $\omega_i$ are the phase and natural frequency of oscillator $i$, respectively.
This model presents a second-order phase transition from the asynchronous to the synchronous regime when we have non-identical oscillators and their frequency is not correlated with the structural properties of the network.

We have considered a uniform initial condition, and the natural frequencies have been obtained from a Lorentz distribution whose probability density function is
\begin{equation}
 f(x; x_0,\gamma) = \frac{1}{\pi} \left[ \gamma \over (x - x_0)^2 + \gamma^2 \right] ,
\end{equation}
where $\gamma$ is the scaling parameter and $x_0$ is the median, here set to zero.
For visualization purposes, we used $\gamma = 10$ in our experiments.
We also considered an Erd\"os -- Renyi network with $N = 250$ nodes, $p = 0.05$, and the coupling $\lambda = 4.5$, which is in the synchronous regime.
Despite these choices, the results would be the same if we were in the subcritical regime or at the critical point.

\section{Stochastic dynamics}

\subsection{Exact solution for an arbitrary stochastic dynamics}
\label{sec:exact_general}

Restricting ourselves to continuous-time Markov chain (CTMC) processes, and assuming we have $N$ agents, each agent can be in one of $K$ mutually exclusive states. Thus, this process will have $M = K^N$ possible states. Here, we use the letters $i$ and $j$ to index all possible microstates. We also define the microstate $i$ by the vector $\bm x_i$, whose components $[\bm x_i]_k \in \{ 0, 1,\cdots, K-1 \}$ denote the state of node $k$. Furthermore, $\bm s(t) = \left[ s_0(t), ... s_{M-1}(t) \right]$ accounts for the probability of having each microstate at time $t$.
The exact probabilities of a process described by its infinitesimal generator $Q$, which is a semi negative definite operator, and can be written as
\begin{equation} \label{eq:exact}
 \dfrac{d \bm s^T (t)}{dt} = \bm s^T (t) Q,
\end{equation}
which solves as
\begin{equation} \label{eq:solved_exact}
 \bm s^T (t) = \bm s^T (0) \exp \left( Q \right).
\end{equation}
Furthermore, as proposed in~\cite{VanMieghem2014}, we can decompose $Q$ and obtain
\begin{equation}
 \bm s^T(t) = \bm s^T(0) \left( \sum_{j=1}^s \Biggl\{ \sum_{k=1}^{m_j} Z_{jk} t^{k-1} \Biggl\} \exp \left( \mu_j t \right) \right),
\end{equation}
where $Z_{jk}$ are independent coefficient matrices, $\mu_j \in \mathbb{C}$ is an eigenvalue of $Q$ having $s$ different eigenvalues, each with multiplicity $m_j$, so that $\sum_{j=1}^s m_j = |K|^N$.
Note also that the multiplicity of $\mu_s = 0$ is the number of absorbing states in the process.

\subsection{Exact solution for the SIS}
\label{sec:exact}

The SIS model is defined as a set of independent Poisson processes modeling both the spread from an infected to a susceptible agent at a rate through contacts $\beta$ and the recovery from infected to susceptible at a rate $\delta$. The contacts are encoded in a graph whose adjacency matrix is denoted $A$.
Following the exact formulation proposed in~\cite{Mieghem2009} and also studied in~\cite{VanMieghem2014, Arruda2018}, we define the state vector in reverse order, i.e, $\bm x_i = \left[ x_N, x_{N-1}, \dots, x_1 \right]$, where $x_k$ is the state at node $k$, $i$ indexes all microstates and is the integer representation of the binary number $\bm x_i$, $i = \sum_{k=1}^N x_k 2^{k-1}$.
The infinitesimal generator that defines the SIS is given by~\cite{Mieghem2009, VanMieghem2014, Arruda2018}
\begin{equation*}
 Q_{ij} =
 \begin{cases}
   \delta &\text{if}  \hspace{1mm} i = j + 2^{m-1}; \\
    & m= 1,2,\dots N; [\bm x_i]_m =1\\
   \beta \sum_k^N A_{mk} [\bm x_i]_k &\text{if}  \hspace{1mm} i = j + 2^{m-1}; \\
    & m= 1,2,\dots N; [\bm x_i]_m =0\\
   - \sum_k Q_{kj} &\text{if} \hspace{1mm} i = j\\
   0 &\text{otherwise}
 \end{cases}
\end{equation*}
which can be solved with Eq.~\ref{eq:exact}, as in Sec.~\ref{sec:exact_general}.

Often, we are interested in analyzing the expected number of in a given state. In this case, we can define a matrix $M$ that aggregates the desired states, allowing us to define macrostate variables as
\begin{equation} \label{eq:rho}
 \rho(t) = \bm s^T(t) M u \frac{1}{N},
\end{equation}
where $u$ is the all-one vector with the appropriate dimension, the product $\bm s^T(t) M$ gives the probability of each macrostate, and the product by $\frac{u}{N}$ takes the average.

\subsection{Cellular automata formulation for the SIS process}

We can use the definition of Poisson processes to understand the cellular automata formulation.
The probability that a susceptible node is infected after a time window of $\Delta t$ is $1 - e^{-\beta n_i \Delta t}$, where $n_i$ is the number of infected individuals in the neighborhood of the node under analysis.
This expression is derived from the superposition theorem and the properties of Poisson processes.
On the other hand, in the case of the sampled Markov chain, this probability is $(1 - \beta \Delta t )^{n_i}$.
The differences between these two models have been studied in~\cite{Gleeson2016}, where the inaccuracies between discrete-time and continuous-time formulations were attributed to the differences between these two formulas.
We emphasize that our arguments are consistent with Ref.~\cite{Gleeson2016}.

In a cellular automaton, the transition between two microstates $i$ and $j$ is the product of the change probabilities of each node state. Formally, using the same indexing scheme as in the previous section, the transition probability matrix is expressed as
\begin{equation}
   P^{CA}_{ij} = \prod_{k=1}^{N} f_k(\bm x_i, \bm x_j),
\end{equation}
where $f_k(\bm x_i(t), \bm x_j(t))$ is a function that determines the probability that node $k$, in state $[\bm x_i]_k$, transitions to state $[\bm x_j]_k$ after a $\Delta t$ step. The product takes into account the probabilities of all events occurring within the same time window.
Thus, the synchronous CA representing the generator formalized for the SIS in the previous section is summarized by $f_k(\bm x_i, \bm x_j)$, given as
\begin{align*}
\scalefont{0.9}
 \begin{cases}
       (1 - \beta \Delta t)^{n_k^i}  &\text{if} \hspace{1mm} [\bm x_i]_k = 0; \hspace{1mm} [\bm x_j]_k = 0 \\
       \delta \Delta t (1 - \beta \Delta t)^{n_k^i} &\text{if} \hspace{1mm} [\bm x_i]_k = 1 \hspace{1mm}; [\bm x_j]_k = 0 \\
       (1 - (1 - \beta \Delta t)^{n_k^i}) &\text{if} \hspace{1mm} [\bm x_i]_k = 0; \hspace{1mm} [\bm x_j]_k = 1 \\
       (1 - \delta \Delta t) + \delta \Delta t (1 - (1 - \beta \Delta t)^{n_k^i}) &\text{if} \hspace{1mm} [\bm x_i]_k = 1; \hspace{1mm} [\bm x_j]_k = 1 \\
 \end{cases},
\end{align*}
where $n_k^i$ is the number of infected neighbors that node $k$ has when we are in the $i$-th microstate. The function $f_k(\bm x_i, \bm x_j)$ is defined under the assumption that an infected node can recover and be reinfected within a $\Delta t$ time window. This approach was proposed in~\cite{Gomez2010}. From a practical point of view, this seems unrealistic. Mathematically, however, we neglect the possibility of longer sequences of recovery and reinfection. This limits the range in which $\Delta t$ can be extended. The order parameter (fraction of infected nodes) can be extracted as $\rho(t) = \bm s^T(t) M u \frac{1}{N}$, where $M \in \{0, 1\}^{2^N \times N}$ contains all possible microstates in bit-reversed order (see Sec.~\ref{sec:exact}, and also~\cite{Mieghem2009} and~\cite{VanMieghem2014} Section 17.2 for a similar formulation).

\subsection{Numerical methods}
\label{sec:numerical}

We use spectral decomposition to efficiently solve the time behavior of the exact model and its discrete versions. Considering an arbitrary function of the square matrix $\bm M(t)$, we have that
\begin{equation}
 f(\bm M ) = \bm V f( \Lambda ) \bm V^{-1},
\end{equation}
where $\bm V$ is a matrix whose columns are the eigenvectors of $\bm M$ and $\Lambda$ is a diagonal matrix whose entries are the eigenvalues of $\bm M$.
Since $\Lambda$ is a diagonal matrix, $f(\cdot)$ will be a scalar function of the eigenvalues.

In the CTMC case, we (i) decompose $Q$ once, (ii) compute $\exp(\Lambda t)$ by computing $\Lambda_{ii} = \exp(\mu_i t)$, and (iii) compute the matrix product $\bm V \bm \exp( \Lambda t ) \bm V^{-1}$.
In this way, we decompose $Q$ only once, which reduces the computational cost of the matrix multiplications.
On the other hand, in the DTMC case, we (i) decompose $P^S$ or $P^{CA}$, depending on the process under analysis, (ii) compute $(\Lambda)^n$ by computing $\Lambda_{ii} = (\mu_i)^n$, where $n$ is a positive integer, and (iii) compute the matrix product $\bm V \Lambda^n \bm V^{-1}$.
This algorithm can obtain the $n$-th power without computing the $(n-1)$-th power, allowing us to obtain the temporal behavior of our equations efficiently.

\subsection{Transition matrices for a star: an example}
\label{sec:transitions_star}

\begin{figure*}[t]
   \includegraphics[width=0.975\textwidth]{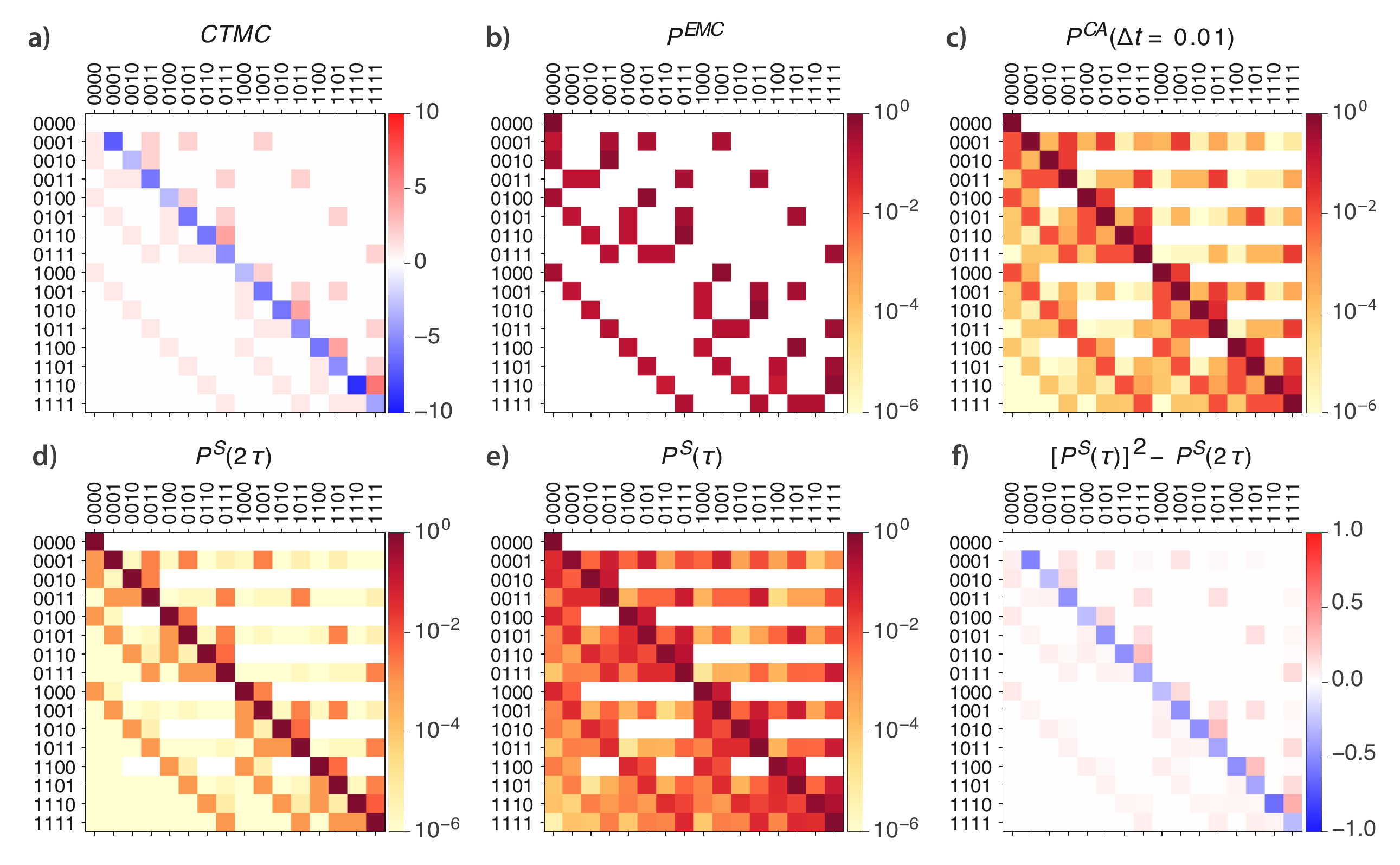}
   \caption{Graphical representation of the infinitesimal generator and different transition probabilities defined for different values of $\delta t$ and methods.
   All matrices were extracted considering a star graph with $N = 5$, $\beta = 2$, and $\delta = 1$.
   In (a), the infinitesimal generator is shown. In (b), the probability transition matrix for the embedded Markov chain. In (c), the cellular automata formulation with $\delta t = 0.01$. In (d) to (f), we illustrate the differences obtained by different sampling frequencies. In (d) with $\Delta t = \tau$, while in (e) with $\Delta t = 2 \tau$, where $\tau = \nicefrac{1}{\max_i |Q_{ii}|}$.
   In (f), we compare the difference in the probability of simulating the process with two steps of $\tau$ versus a single step of $2\tau$.}
   \label{fig:Transitions_Star}
\end{figure*}

Fig.~\ref{fig:Transitions_Star} shows a graphical representation for different matrices representing the SIS in a star with $N = 5$ nodes, $\beta = 2$, and $\delta = 1$.
In Fig.~\ref{fig:Transitions_Star} (a), we show the infinitesimal generator that defines the CTMC, while in (b), we show the probability transition matrix for the embedded Markov chain.
In (c), we show the cellular automata formulation with $\Delta t = 0.01$.
In (d) to (f), we illustrate the differences obtained by different sampling frequencies. In (d) with $\Delta t = \tau$, while in (e) with $\Delta t = 2 \tau$, where $\tau = \nicefrac{1}{\max_i |Q_{ii}|}$.
Finally, in (f), we compare the difference in terms of the probability of simulating the process considering two steps of $\tau$ with a single step of $2\tau$.

\begin{figure}[b!]
   \includegraphics[width=0.98\columnwidth]{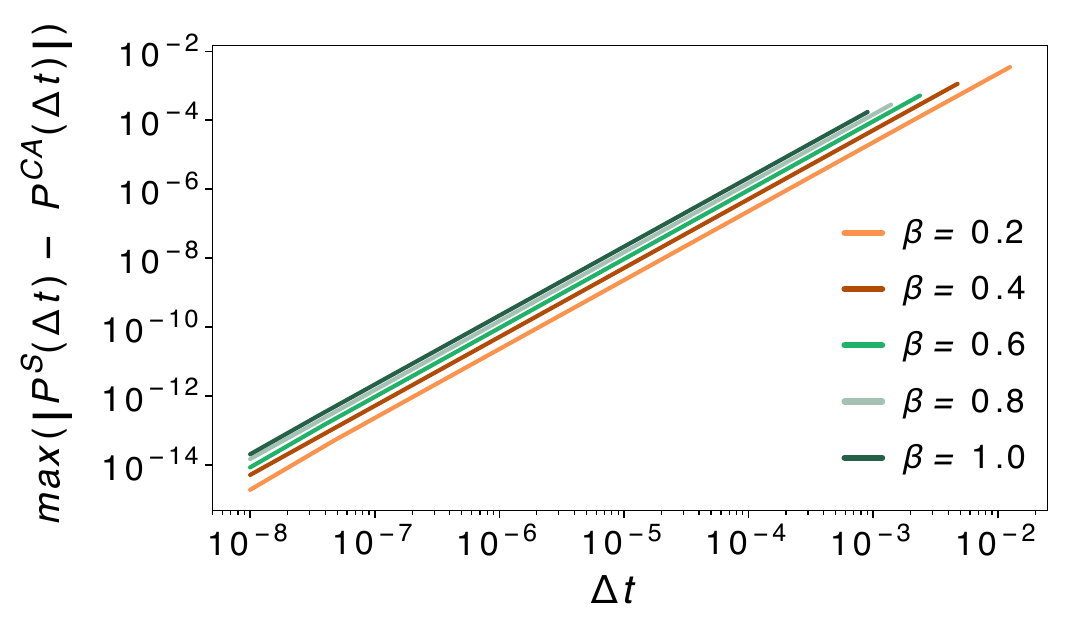}
   \caption{Numerical experiment showing that $\lim_{\Delta t \rightarrow 0} |P^S(\Delta t) - P^{CA}(\Delta t)| = 0$ considering the SIS process with different values of $\beta$, $\delta = 0.5$ on an random network with $N = 10$ nodes and average degree $\langle k \rangle = 5$.}
   \label{fig:CA_Convergence}
\end{figure}

\begin{figure*}[t!]
   \includegraphics[width=0.98\textwidth]{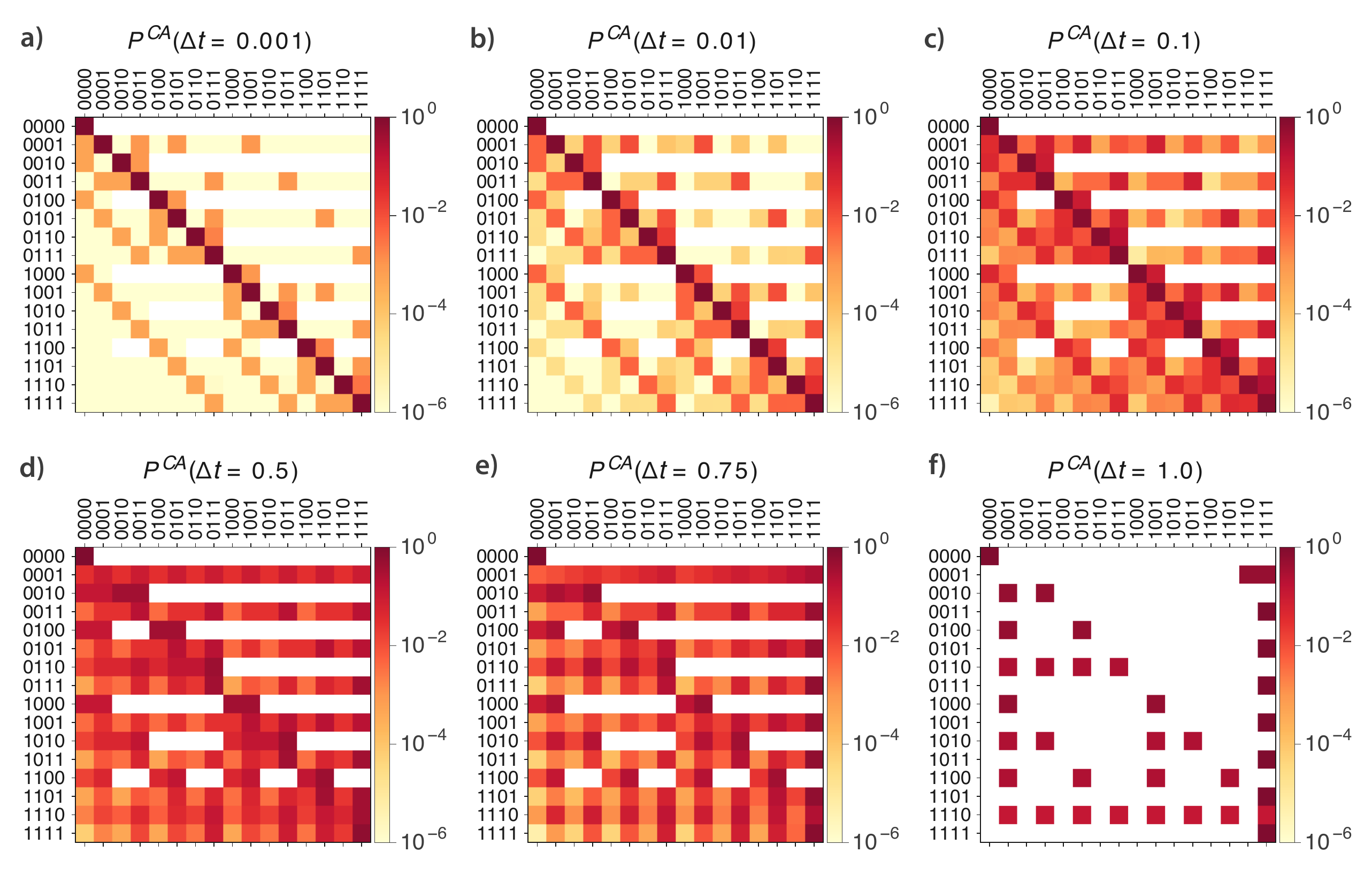}
   \caption{Graphical representation of the probability transition matrix for the cellular automata and different values of $\Delta t$.
   All the matrices were extracted considering a star graph with $N = 5$, $\beta = 1$, and $\delta = 0.5$.
   From (a) to (f), we considered $\Delta t = 0.001, 0.01, 0.1, 0.5, 0.75, 1.0$.}
   \label{fig:CA_Convergency_Star}
\end{figure*}

The EMC shows a similar pattern of transitions as the infinitesimal generator, as can be seen by comparing figures~\ref{fig:Transitions_Star} (a) and (b). In this case, however, the diagonal elements are excluded.
It is also instructive to note that the EMC is similar to a non-lazy random walk on the graph of transitions.
Next, in Fig.~\ref{fig:Transitions_Star} (c), we illustrate an extreme case where $\Delta t \gg \tau$. Compared to the EMC transitions, we can see that the probability transition matrix allows more transitions. This can be observed by comparing the number of non-zero elements in figures~\ref{fig:Transitions_Star} (b) and (c).

Finally, in figures~\ref{fig:Transitions_Star} (d) to (e), we compare the effect of $\Delta t$ in successive steps.
One would hope that, for a fixed set of rates, one step in a model defined with $2 \Delta t$ would yield the same results as two steps in a model defined with $\Delta t$.
In practice, the product of the state vector and the transition probability matrix gives a step in a discrete model.
In Fig.~\ref{fig:Transitions_Star}(e), we can compare the effects of different $\Delta t$'s by comparing the difference between a step in a model with $\Delta t = \tau$ and the same model but with $\Delta t = 2 \tau$, where $\tau = \nicefrac{1}{\max_i |Q_{ii}|}$.
Note that there are visible differences between these two cases. The ideal case would be an all-zero matrix.

\subsection{Cellular automata convergency and the emergence of a second absorbing state}
\label{sec:2_absorbing}

As discussed in the main text, the sampled Markov chain, $P^S$, converges to the CTMC as $\Delta t$ goes to zero.
However, the convergence from the cellular automata formulation to the CTMC is less straightforward. In fact, the cellular automata assumes that we can have at most one reinfection within a $\Delta t$ time window. While this is a reasonable assumption both in practice and in terms of expected probabilities, it is still an approximation.
Therefore, in Fig.~\ref{fig:CA_Convergence}, we provide a numerical experiment that suggests this convergence, where we show the maximum of the absolute difference between $P^S(\Delta t)$ and $P^{CA}(\Delta t)$ as a function of $\Delta t$.
This figure shows that, regardless of $\beta$, the two matrices become more similar as $\Delta t \rightarrow 0$.
In addition, we recall that the probability that the events generated by two Poisson processes occur at the same time is zero. Thus, in the limit of $\Delta t \rightarrow 0$, the approximation that we can have only one reinfection within a $\Delta t$ time window should not be a problem. Note that the probability of reinfection scales with $O((\Delta t)^2)$.

Next, in Fig.~\ref{fig:CA_Convergency_Star}, we evaluate the effect of $\Delta t$ on the probability transitions for the cellular automata, focusing on relatively large values.
From (a) to (e), we have $\Delta t = 0.001, 0.01, 0.1, 0.5, 0.75, 1.0$, respectively.
For visualization purposes, we plot these matrices on a star with $N = 5$ nodes and $\beta = 1$ and $\delta = 0.5$.
Note that despite the different rates, Fig.~\ref{fig:Transitions_Star} (c) and Fig.~\ref{fig:CA_Convergency_Star} (b) are the same because the ratio $\nicefrac{\beta}{\delta}$ is the same.

\subsection{Continuous-time Fourier transform}
\label{sec:ctft}

The Nyquist-Shannon theorem and the uncertainty principle are based on the Fourier transform. Thus, it is instructive to analyze the Fourier transform of the SIS exact solution, Eq.~\eqref{eq:solved_exact}.
Denoting the Fourier transform of such a function as $\bm S^T(\omega) = \mathcal{F}_t \{\bm s^T \} (\omega)$, we have
\begin{equation} \label{eq:Somega}
 \bm S^T(\omega) = \bm s^T(0) \left( \sum_{j=1}^s \sum_{k=1}^{m_j} Z_{sk} g(\mu_s, k) \right),
\end{equation}
where $g(\mu_s, k)$ is a function of the eigenvalue $\mu_s$ and its multiplicity $k$, defined as
\begin{equation}
 g(\mu_s, k) =
 \begin{cases}
    \pi \delta(\omega) + \frac{1}{\mathrm{i} \omega} \hspace{3mm} & \text{if} \hspace{3mm} \mu_s = 0 \\
    \frac{(k-1)!}{(\mu_s + \mathrm{i} \omega)^{k} } \hspace{3mm} &\text{if} \hspace{3mm} \mu_s < 0
 \end{cases}
\end{equation}

Due to the linearity property of the Fourier transform, we have that the frequency spectrum of the order parameter $\rho$, Eq.~\eqref{eq:rho}, is obtained by weighting the frequencies in Eq.~\eqref{eq:Somega}.
The magnitude of the frequency spectrum for each exponential will have a peak at zero whose height will be $| \mu_s|$.
Note that if the process has at least one absorbing state, then $a = 0$, and $c$ is a constant that depends on the initial condition and the spectral properties of $Q$, i.e., the coefficients $Z_{sk}$ and the eigenvalue multiplicity.
Moreover, since our process consists of a summation of exponential functions, the magnitude of the spectrum is greater than zero at any frequency but will decrease as $\frac{c}{\sqrt{a^2 + \omega^2}}$, where $a = \min_s \mu_s$.
This observation highlights a fundamental limitation of sampled versions of the process defined in Eq.~\eqref{eq:exact}. Since any sampling also implies a fixed frequency bandwidth $[-\omega_{\max}, \omega_{\max}]$, any frequency $\omega > \omega_{\max}$ will be neglected, implying aliasing.
In this case, using the inverse Fourier transform with a limited signal
\begin{equation}
 \hat{\bm s} (t) = \int_{-\omega_{\max}}^{\omega_{\max}} \bm S^T(\omega) e^{\mathrm{i} \omega t} d\omega,
\end{equation}
similar to the proof of the Nyquist-Shannon theorem in~\cite{Shannon1949}.

\end{document}